\documentclass[galaxies,article,submit,moreauthors,pdftex]{Definitions/mdpi} 

\firstpage{1} 
\pubvolume{1}
\issuenum{1}
\articlenumber{0}
\pubyear{2021}
\copyrightyear{2020}
\datereceived{} 
\dateaccepted{} 
\datepublished{} 
\hreflink{https://doi.org/} % If needed use \linebreak
\pdfoutput=1

\Title{Bent it like FRs: extended radio AGN in the COSMOS field and their large-scale environment}

\TitleCitation{Bent it like FRs in COSMOS}

\Author{Eleni Vardoulaki$^{1,*}$\orcidA{}, Franco Vazza$^{2,5,6}$\orcidB{}, Eric F. Jim\'{e}nez-Andrade$^{3}$\orcidC{}, Ghassem Gozaliasl$^{4}$\orcidD{}, Alexis Finoguenov$^{4}$\orcidE{}, and Denis Wittor$^{5,6}$\orcidF{}}

\AuthorNames{Eleni Vardoulaki, Franco Vazza, Eric F. Jim\'{e}nez-Andrade,  Ghassem Gozaliasl, Alexis Finoguenov, and Denis Wittor}

\AuthorCitation{Vardoulaki, E.; Vazza, F.; Jim\'{e}nez-Andrade, E. F.; Gozaliasl, G.; Finoguenov, A., and Wittor, D.}
\address{%
$^{1}$ \quad Th\"{u}ringer Landessternwarte, Sternwarte 5, 07778 Tautenburg, Germany\\
$^{2}$ \quad Physics and Astronomy Department, University of Bologna, via Gobetti 93/2, Bologna, 40122, Italy \\
$^{3}$ \quad National Radio Astronomy Observatory, 520 Edgemont Road, Charlottesville, VA 22903, USA \\
$^{4}$ \quad Department of Physics, University of Helsinki, P. O. Box 64, FI-00014, Helsinki, Finland \\
$^{5}$ \quad Hamburger Sternwarte, Gojenbergsweg 112, 21029 Hamburg, Germany \\
$^{6}$ \quad Dipartimento di Fisica e Astronomia, Universita di Bologna, Via Gobetti 93/2, 40122, Bologna, Italy\\}

\corres{Correspondence: elenivard@gmail.com}% Tel.: (optional; include country code; if there are multiple corresponding authors, add author initials) +xx-xxxx-xxx-xxxx (F.L.)}

\abstract{One of the fascinating topics in radio astronomy is how to associate the complexity of observed radio structures to their environment, in order to understand their interplay and the reason for the plethora of radio structures found in surveys. In this project, we explore the distortion of the radio structure of Fanaroff-Riley (FR) type radio sources in the VLA-COSMOS Large Project at 3 GHz, and relate it to their large-scale environment. We quantify the distortion by using the angle formed between the jets/lobes of two-sided FRs, namely bent angle (BA). Our sample includes 108 objects in the redshift range $0.08 < z < 3$, which we cross-correlate to a wide range of large-scale environments (X-ray galaxy groups, density fields, and cosmic web probes) in the COSMOS field. The median BA of FRs in COSMOS at $z_{\rm med} \sim 0.9$ is 167.5$^{+11.5}_{-37.5}$ degrees. We do not find significant correlations between BA and large-scale environments within COSMOS, covering scales from a few kpc to several  hundred Mpc, nor between BA and host properties. Finally, we compare our observational data to magnetohydrodynamical (MHD) adaptive-mesh simulations ENZO-MHD of two FR sources at $z$ = 0.5 and at $z$ = 1. Although the scatter in BA of the observed data is large, we see an agreement between observations and simulations in the bent angles of FRs, following a mild redshift evolution with BA. We conclude that the dominant mechanism affecting the radio structures of FRs could be the evolution of the ambient medium, where higher densities and longer depths at lower redshifts allow for more space for jet interactions. }


\begin{document}
%%%%%%%%%%%%%%%%%%%%%%%%%%%%%%%%%%%%%%%%%%

\section{Introduction}

Extended radio active galactic nuclei (AGN) often deviate from a straight radio structure and their shapes present a plethora of distortions. Past and current radio surveys add on the complexity of radio structures, especially high sensitivity and high resolution surveys since they reveal features in the radio structure not seen with past surveys. The reason behind bent radio structures is likely to be a complex phenomenon. Relative motions of the source or of the jets through a dense intergalactic (IGM) or intracluster medium (ICM), as well as interaction with nearby sources, can cause deviations from the expected straight radio structure \citep[e.g.][]{smolcic07, garon19}. Thus the role of the large-scale environment is considered crucial in shaping extended radio AGN. The degree of complexity varies from source to source, and performing a study of different types of environments can help us identify the reason why extended radio AGN get their jets bent.

In this study we examine the degree of distortion of radio AGN in the COSMOS field \citep{scoville07} in order to take advantage of the plethora of multi-wavelength data and environmental probes. We quantify this distortion by the angle jets/lobes form to each other, namely bent angle (BA). We compare to large-scale environments in COSMOS, and in particular to the X-ray galaxy groups studied by Gozaliasl et al. \citep{gozaliasl19}, the density fields studied by Scoville et al. \citep{scoville13}, and the cosmic-web probes studied by Darvish et al. \cite{darvish15, darvish17}. To our knowledge, our study of BA in relation to galaxy groups is the first one of its kind. The redshift range we cover is $0.08 < z < 3$ and we examine environmental probes from a few kpc to 500 Mpc. Finally, we compare to two simulated radio sources at redshifts $z = 0.5$ and $z = 1$, generated by  magnetohydrodynamical  (MHD) adaptive-mesh simulations ENZO-MHD by Vazza et al. \citep{vazza21}.
 
%%%%%%%%%%%%%%%%%%%%%%%%%%%%%%%%%%%%%%%%%%
\section{Sample}
\label{sec:sample}

\begin{figure}[!ht]
  \resizebox{\hsize}{!}
 {\includegraphics{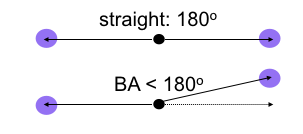}}
 \resizebox{\hsize}{!}
{\includegraphics[ width=\linewidth, trim={7cm 5cm 6.5cm 4.993cm},clip]{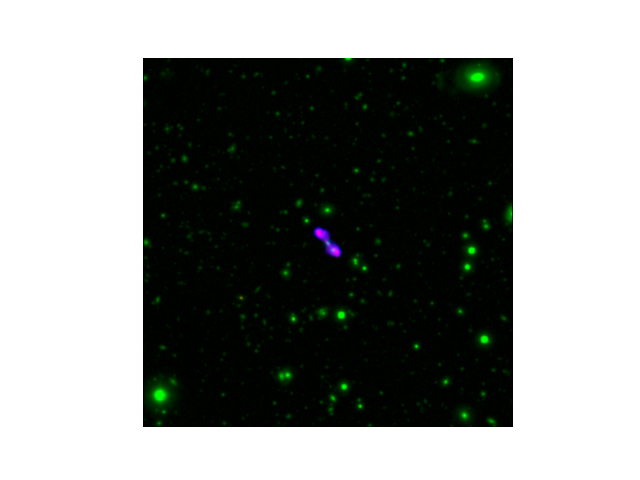}%trim={<left> <lower> <right> <upper>}
 \includegraphics[width=\linewidth, trim={7cm 5.5cm 5.5cm 3.5cm},clip]{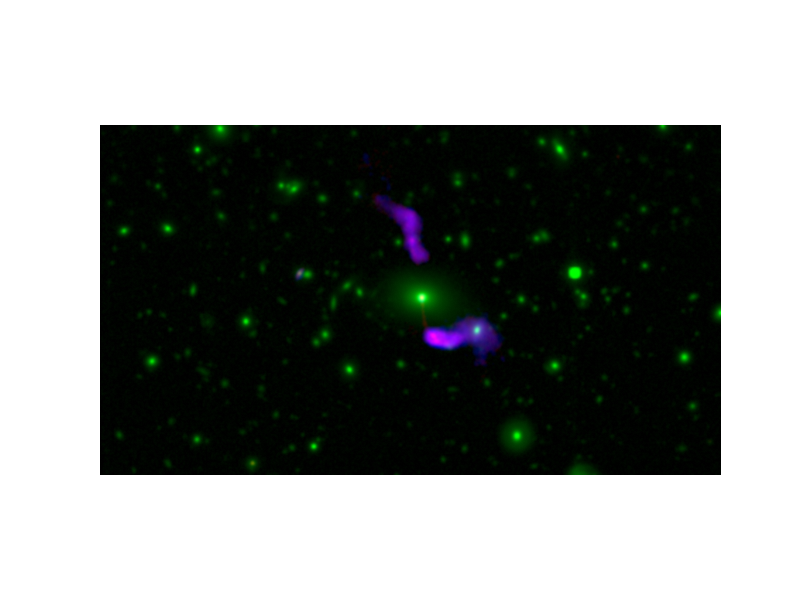}
\includegraphics[width=\linewidth, trim={5.5cm 2.5cm 5.5cm 5.5cm},clip]{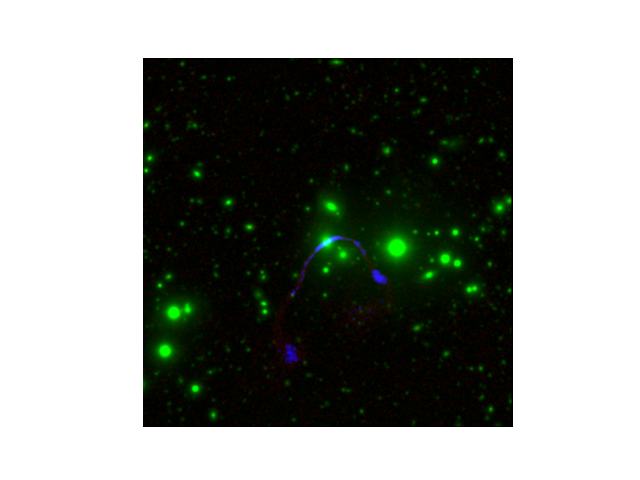}
            }
                 
       \caption{Top: Sketch showing how the bent angle was measured. Bottom: Examples of bent FRs from \cite{vardoulaki21}, for BA = 180$^{\circ}$ (left), 139$^{\circ}$ (middle), and 57$^{\circ}$ (right). The radio emission at 1.4 \& 3 GHz is shown in purple, and the infrared Ultra-VISTA stacked mosaic with green. The latter identifies the host of the radio AGN.
   }
              \label{fig:bent_sketch}%
    \end{figure}

The sample is drawn from Vardoulaki et al. \citep{vardoulaki21}, who classified the extended radio AGN at 3 GHz VLA-COSMOS based on the FR-type classification scheme \citep{fr74} as FRII or edge brightened, FRI or edge darkened, and FRI/FRII or hybrids; the latter have a jet on one side an a lobe on the other. From the 130 FRs reported in \cite{vardoulaki21} we use 108 for the current analysis where bent angles (BA) are available. These were measured through visual inspection on the projected sources, which could add a source of uncertainty. The values can be found in \cite{vardoulaki21}. The BA is given in degrees and is the angle formed between jets/lobes with respect to each other in a two-sided source (Fig.~\ref{fig:bent_sketch}). The Vardoulaki et al. \cite{vardoulaki21} sample also includes jet-less radio AGN and star-forming galaxies which we do not include in this analysis due to their jet-less radio structure. 

%%%%%%%%%%%%%%%%%%%%%%%%%%%%%%%%%%%%%%%%%%
\section{Analysis and Results}

 \subsection{Bent angle vs FR type}

\begin{figure}[!ht]
  \resizebox{\hsize}{!}
 {\includegraphics[width=0.6cm]{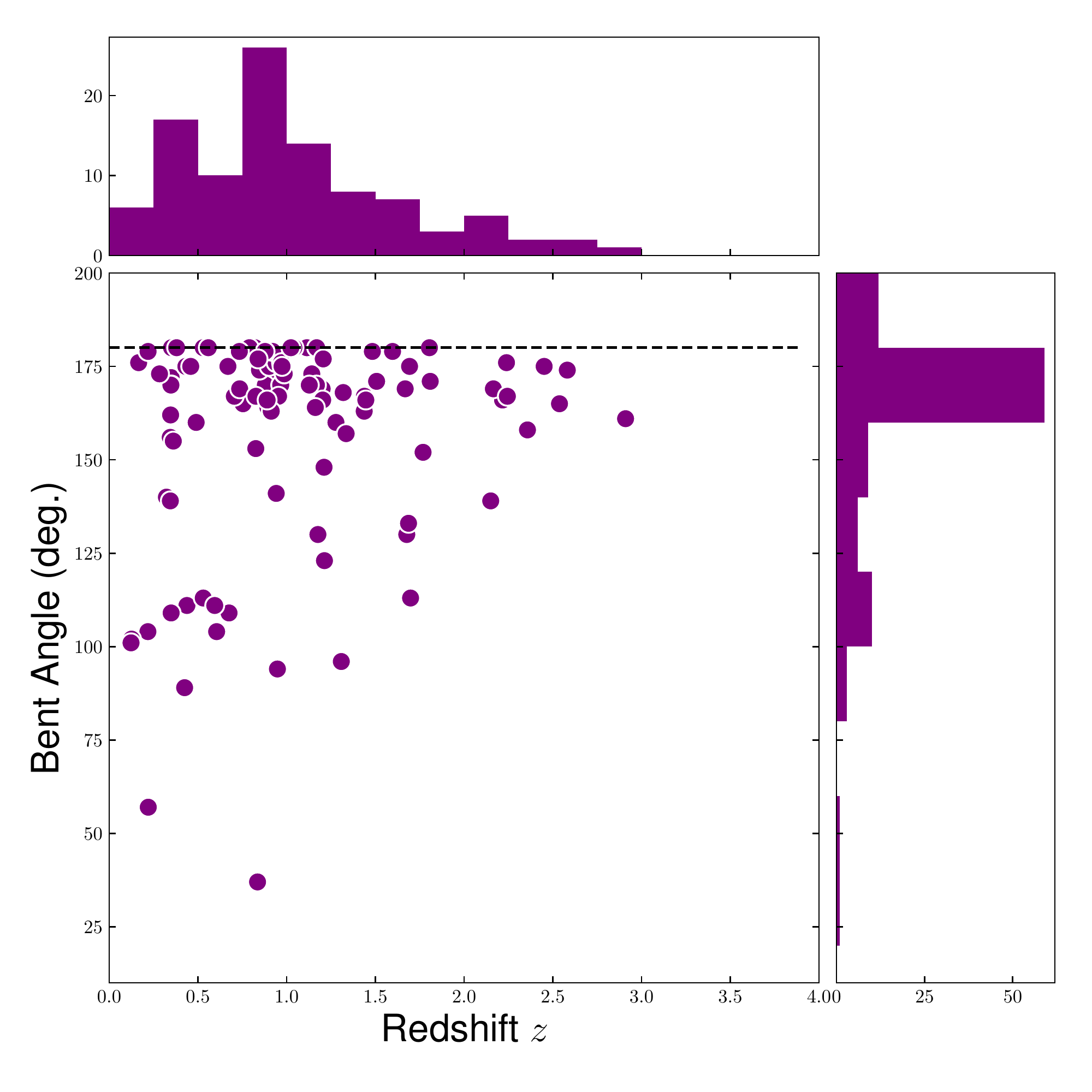}
            }
       \caption{Bent angle in degrees of COSMOS FR sources at 3 GHz versus redshift. For reference, the large-scale environments we use in the analysis cover the redshift range up to $z \sim 3$, depending on the probe.   
   }
              \label{fig:bent_z}%
    \end{figure}

The median values of the 108 FRs with BA are listed in Table~\ref{table:ba_values}. The FR population at 3 GHz VLA-COSMOS has a median BA = 167.5 $\pm$ 16.1 degrees, with minimum BA = 37 degrees and maximum BA = 180 degrees. The latter value represents objects in which their jets/lobes are on a straight line (see Fig.~\ref{fig:bent_sketch}). The median redshift of the FR population presented here is $z_{\rm med}$ = 0.9. In Fig.~\ref{fig:bent_z} we show the redshift range covered by our sample in relation to the BA of the objects. In Table~\ref{table:ba_values} we also provide the median values for the different FR types. FRIs and FRI/FRIIs in our sample are more bent than FRIIs, albeit, there is a large scatter in the values and an overlap in the distributions.

\begin{specialtable}[H] 
\caption{Median bent angle of FRs sources in our sample.}
\label{table:ba_values}
%%% \tablesize{} %% You can specify the fontsize here, e.g., \tablesize{\footnotesize}. If commented out \small will be used.
\begin{tabular}{l c c c c c c}
\toprule
\textbf{Radio} & \textbf{$N$} & \multicolumn{4}{c}{\textbf{Bent angle (degrees)}} \\
\textbf{class} &  & \textbf{median $\pm$ error} & \textbf{median$^{84\%}_{16\%}$} & \textbf{min} & \textbf{max}\\
\midrule
FRII & 59 & 170.0 $\pm$ 22.1 & 170.0$^{179.0}_{140.1}$ & 57.0 & 180.0\\
\\
FRI/FRII & 25 & 165.0 $\pm$ 33.0 & 165.0$^{175.3}_{132.0}$ & 101.0 & 180.0 \\
\\
FRI & 24 & 167.0 $\pm$ 34.0 & 167.0$^{180.0}_{104.0}$ & 37.0 & 180.0 \\
\\
\bottomrule
\end{tabular}
\end{specialtable}

\subsection{Bent angle vs large-scale environment and galaxy type}

\begin{figure}[!ht]
  \resizebox{\hsize}{!}
 {\includegraphics{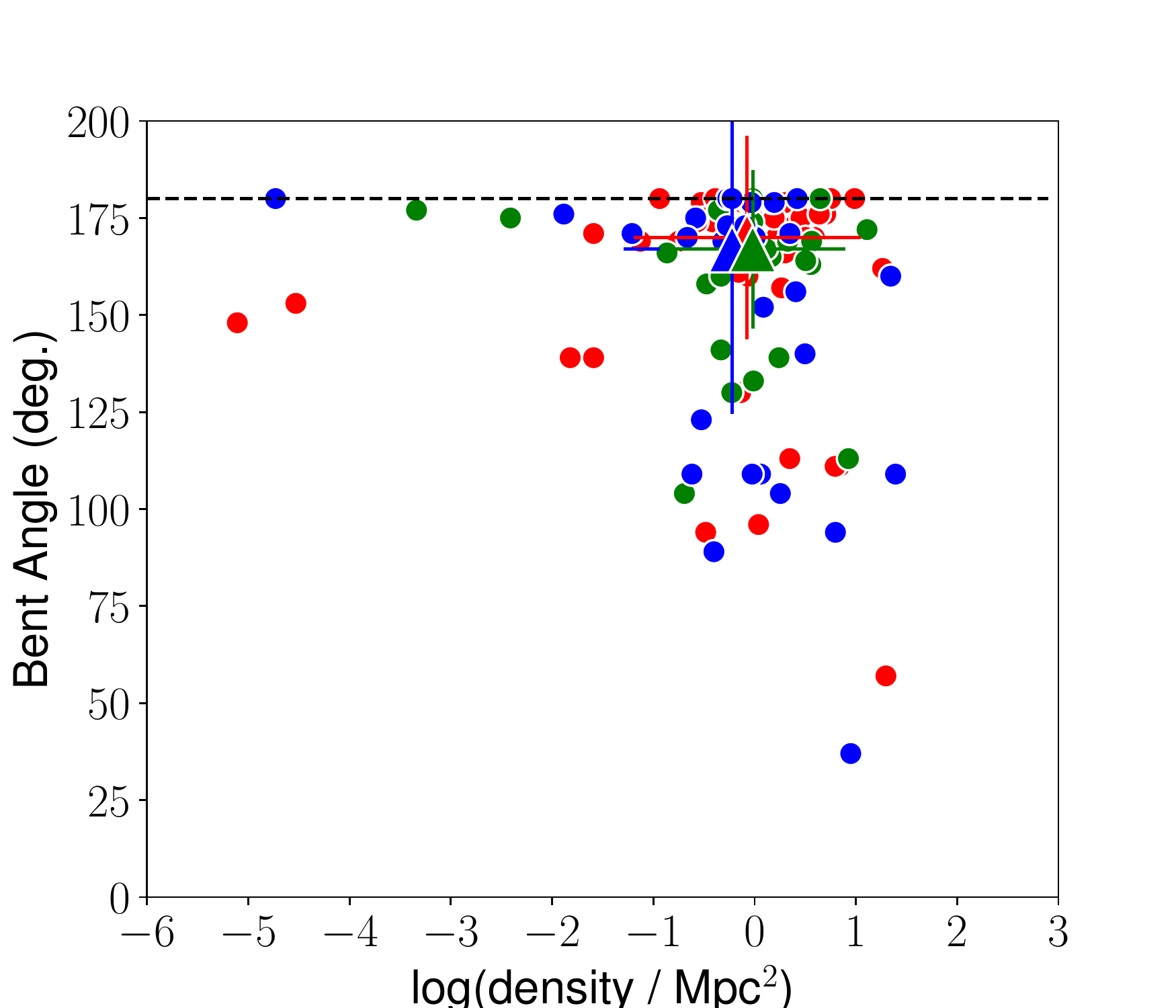}
\includegraphics[width=15cm]{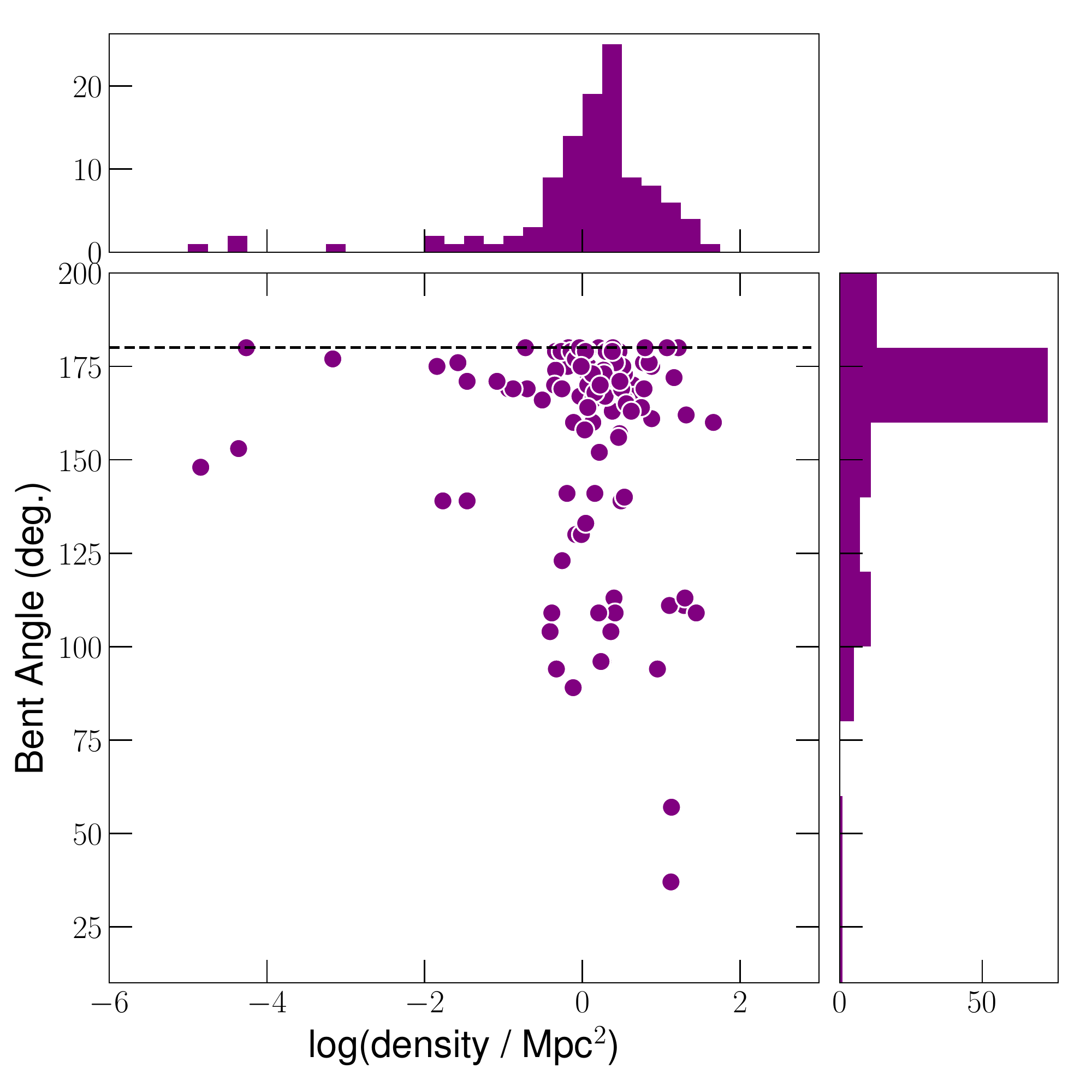}
            }
                 
       \caption{{\bf Left}: Bent angle in degrees for FRIIs (red), FRI/FRIIs (green), and FRIs (blue) with respect to density fields per co-moving Mpc$^{2}$ from \cite{scoville13}. Large triangles are the mean values for FRIIs (red), for FRI/FRIIs (green), and for FRIs (blue), while the error bar gives the standard deviation. 
       {\bf Right}: Same plot as on the left panel, will all FRs plotted as purple circles. In both panels, the horizontal dashed line marks a straight radio source where the jets or lobes form 180$^{\circ}$ angle.
   }
              \label{fig:bent_dens}%
    \end{figure}

We firstly investigate the BA of FRs in relation to large-scale environments of the order of Mpc scales. We cross-correlate the FR sample to the density fields in COSMOS \citep{scoville13} to infer relations to BA. In particular we make use of the density per co-moving Mpc$^{2}$ calculated using adaptive smoothing and Voronoi tessellation \cite{scoville13}. The density fields are given in a cube of redshift slices\footnote{Data are publicly available at http://irsa.ipac.caltech.edu/data/COSMOS/ancillary/densities/} up to redshift of 3, which includes all the objects in our sample. In Fig.~\ref{fig:bent_dens} we plot the bent angle parameter for our FRs with respect to the density per Mpc$^{2}$. Most of the FRs in our sample lie in low-density environments, with median number density values of 0.84 $\pm$ 0.11 Mpc$^{-2}$ for FRIIs, 0.95 $\pm$ 0.18 Mpc$^{-2}$ for FRI/FRIIs, and 0.59 $\pm$ 0.10 Mpc$^{-2}$ for FRIs. We do not find a correlation between BA and COSMOS density fields. Objects with small BA ($<100^{\circ}$) can be found in similar density environments as the ones above that angle, namely with densities of $\sim$10 per Mpc$^{2}$. We do not find objects at 3 GHz VLA-COSMOS with small BA ($<75^{\circ}$) in less dense environments ($<$ 10 Mpc$^{-2}$).

\begin{figure}[!ht]
  \resizebox{\hsize}{!}
 {\includegraphics{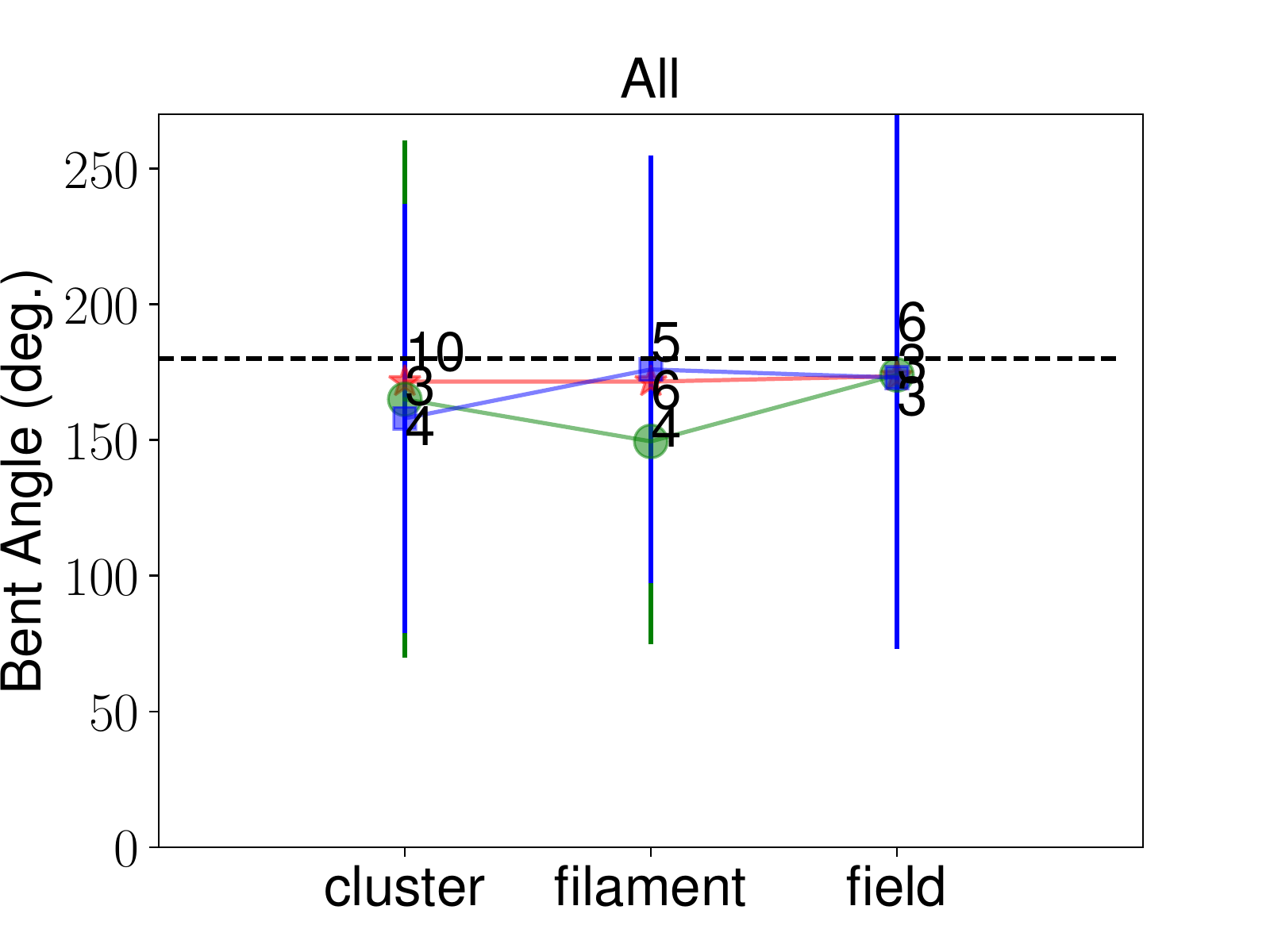}
 \includegraphics{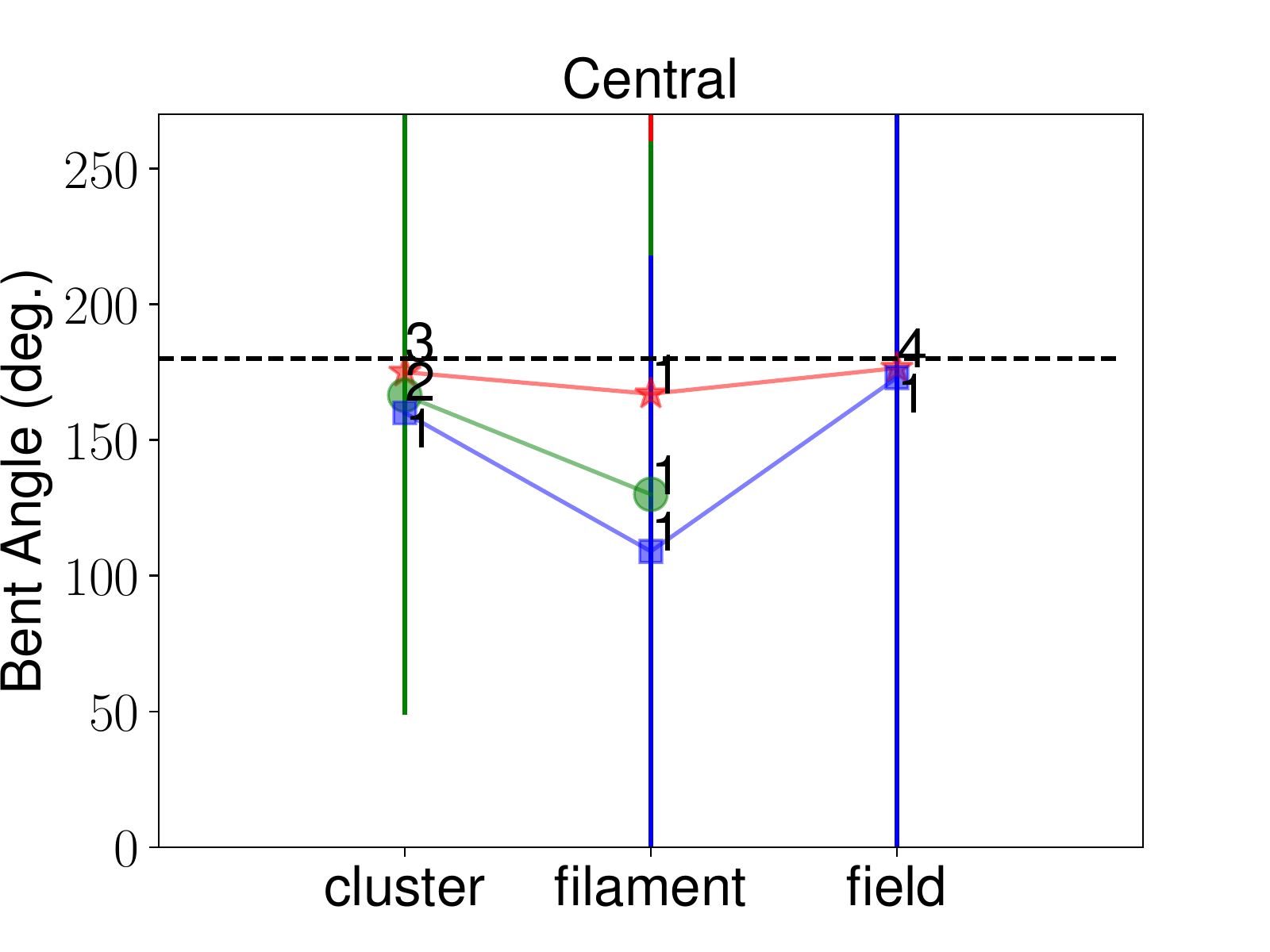}
            }
              \resizebox{\hsize}{!}
 {\includegraphics{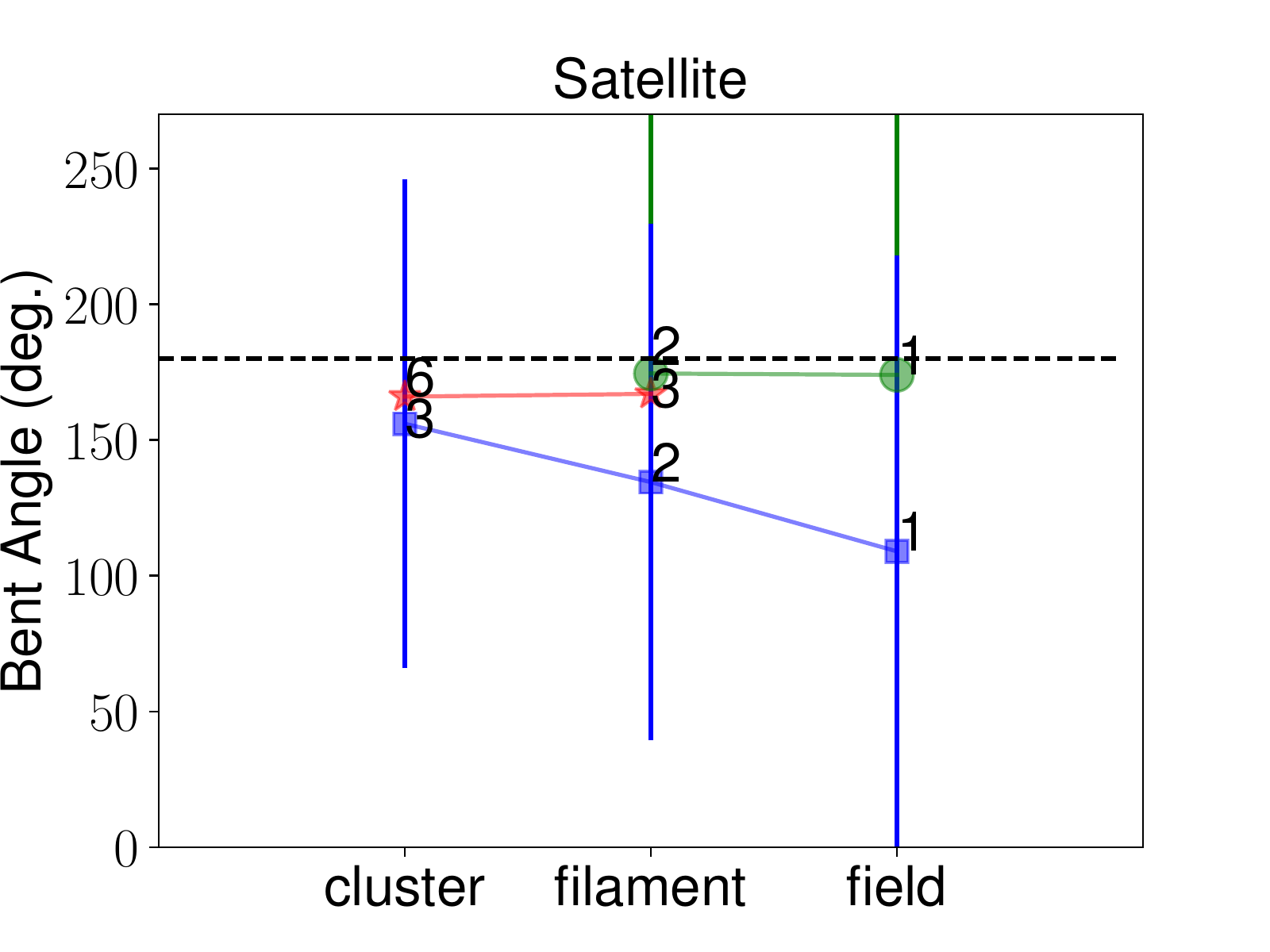}
 \includegraphics{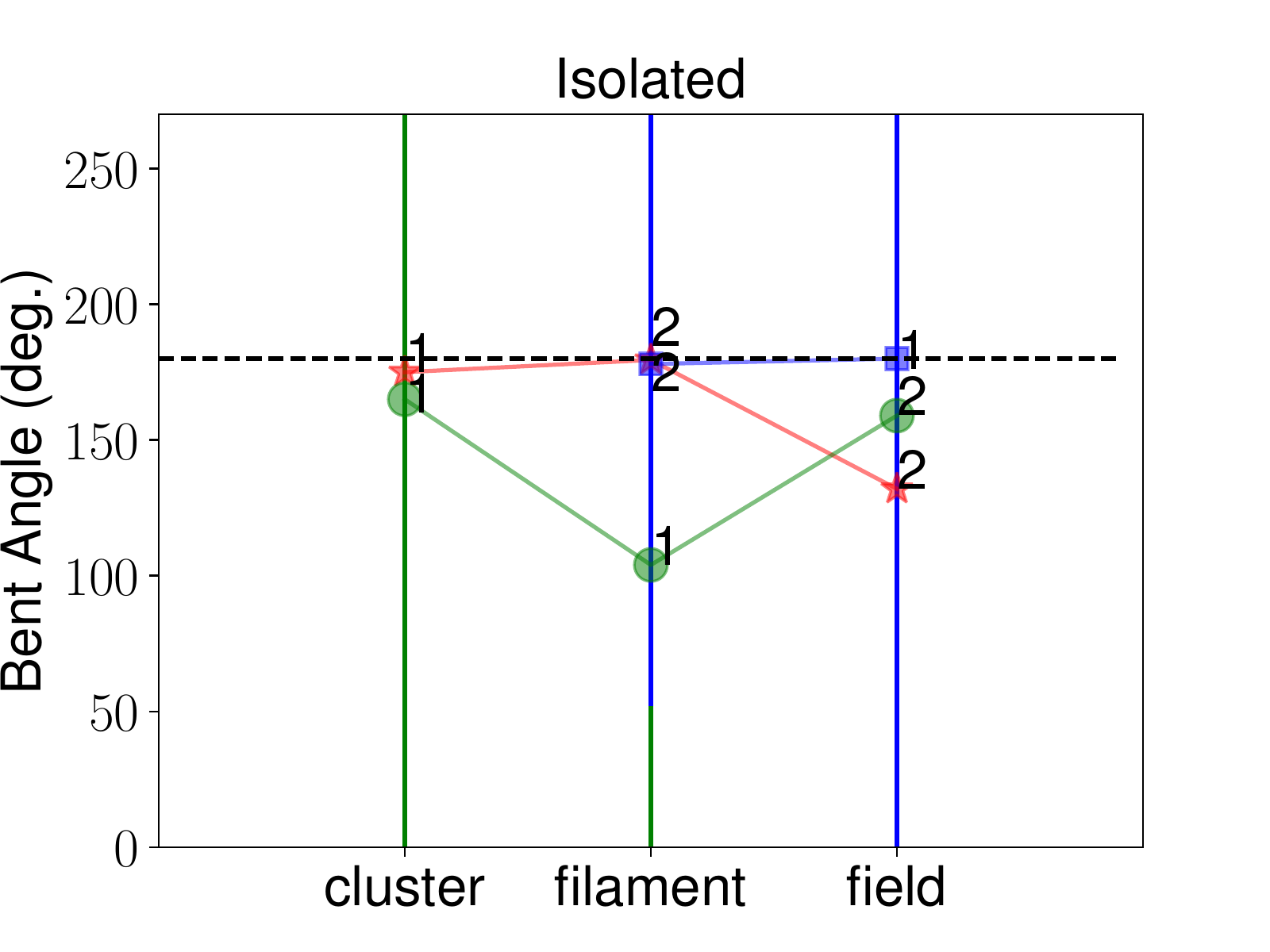}
            }
                 
       \caption{Bent angle in degrees for FRIIs (red), FRI/FRIIs (green), and FRIs (blue) with respect to environment (cluster, filament, and field) depending on galaxy type (central, satellite, isolated), as defined in \cite{darvish17}. The horizontal dashed line marks a straight radio source where the jets or lobes form 180$^{\circ}$ angle. The numbers next to the symbols show the number of objects in that bin. The errors are the errors of the median ($\sim$ median / $\sqrt{N}$).
   }
              \label{fig:bent_darv}%
    \end{figure}

In Fig.~\ref{fig:bent_darv} we compare the median bent angle of the FRII, FRI/FRII, and FRI objects in our sample to the large-scale environments as characterised by the study of \citep{darvish17} up to redshift of 1.2. Darvish et al. \citep{darvish17} characterised the cosmic web environment as cluster, filament, or field, and host environment as central, satellite, or isolated. In Table~\ref{tab:darvish_env_stats} we list the number of FR objects in each cosmic web environment.

\begin{specialtable}[H] 
\caption{FRs with BA measurements in cosmic web probes. Cross-correlation of the FR and COM AGN samples with \cite{darvish17} with respect to different environments (cluster, filament, field). In the parentheses we give the percentage over the total cross-matched number of objects with the same type (within these environments).}
\label{tab:darvish_env_stats}
%%% \tablesize{} %% You can specify the fontsize here, e.g., \tablesize{\footnotesize}. If commented out \small will be used.
\begin{tabular}{l l l l l l l l l}
\toprule
\multicolumn{1}{c}{\textbf{Radio class}}   &      \multicolumn{1}{c}{\textbf{cluster}}  &  \multicolumn{1}{c}{\textbf{filament}}   &  \multicolumn{1}{c}{\textbf{field}}                \\     
\midrule
FRII         & 10 (46\%)& 6 (27\%) & 6 (27\%) \\
FRI/FRII  & 4 (36\%) &  4 (36\%) & 3 (28\%) \\
FRI          & 4 (27\%) & 8 (53\%)& 3 (20\%)\\
\bottomrule
\end{tabular}
\end{specialtable}

In Fig.~\ref{fig:bent_darv} we do not observe a clear relation between BA and cosmic web probes, probably due to the small number statistics (Table~\ref{tab:darvish_env_stats}). Only $\sim$ 44\% of our FR sample  is associated with the cosmic web probes characterised by Darvish et al., and mostly sources with BA $>$ 150 degrees. From Fig.~\ref{fig:bent_darv} we see an indication that FRI/FRIIs in filaments are more bent than FRIs and FRIIs, but the large scatter affects our interpretation. Similarly, FRIs in satellite hosts are more bent in filaments, with median BA below 150 degrees.

\subsection{Bent angle vs X-ray galaxy group properties}

\begin{figure}[!ht]
  \resizebox{\hsize}{!}
 {\includegraphics{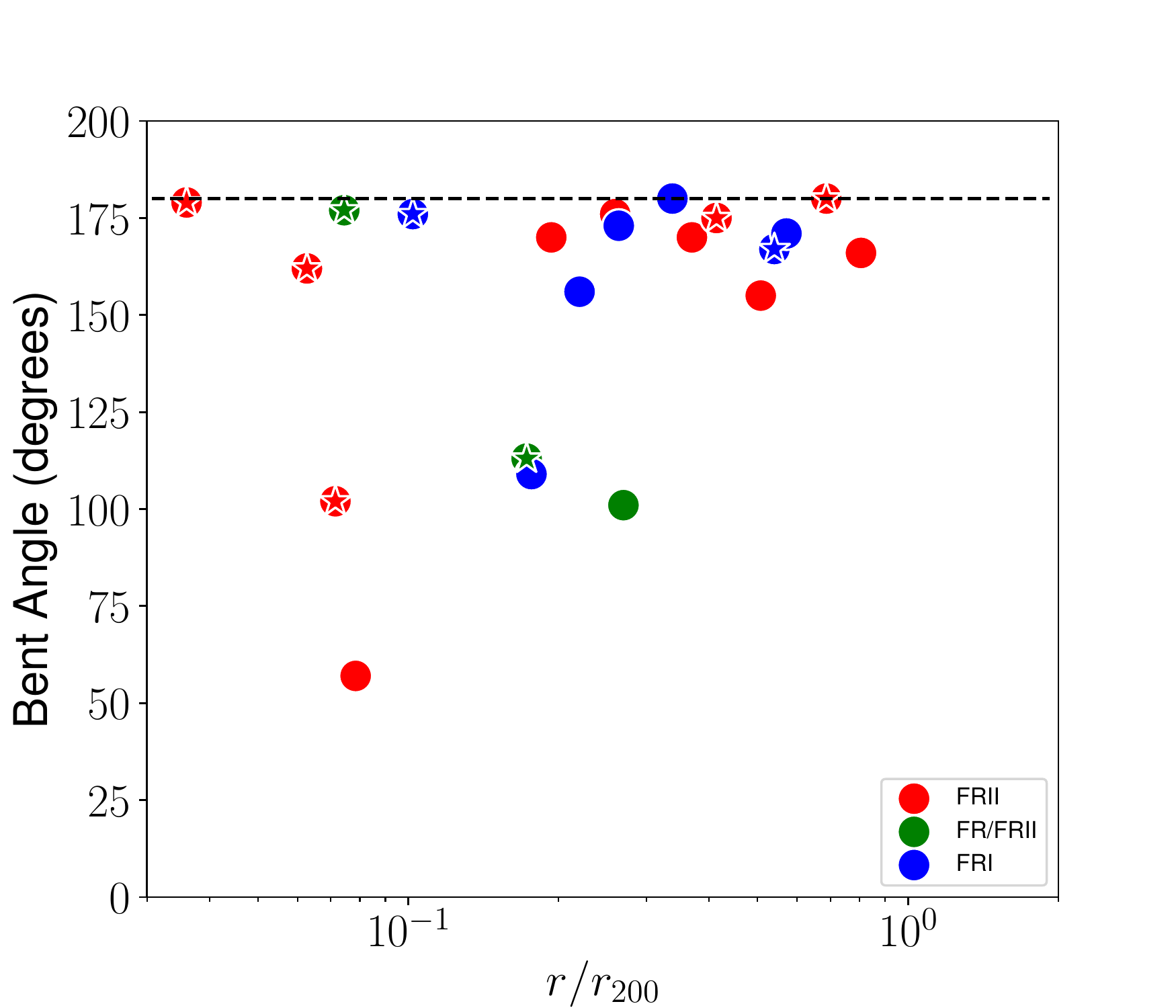}
\includegraphics{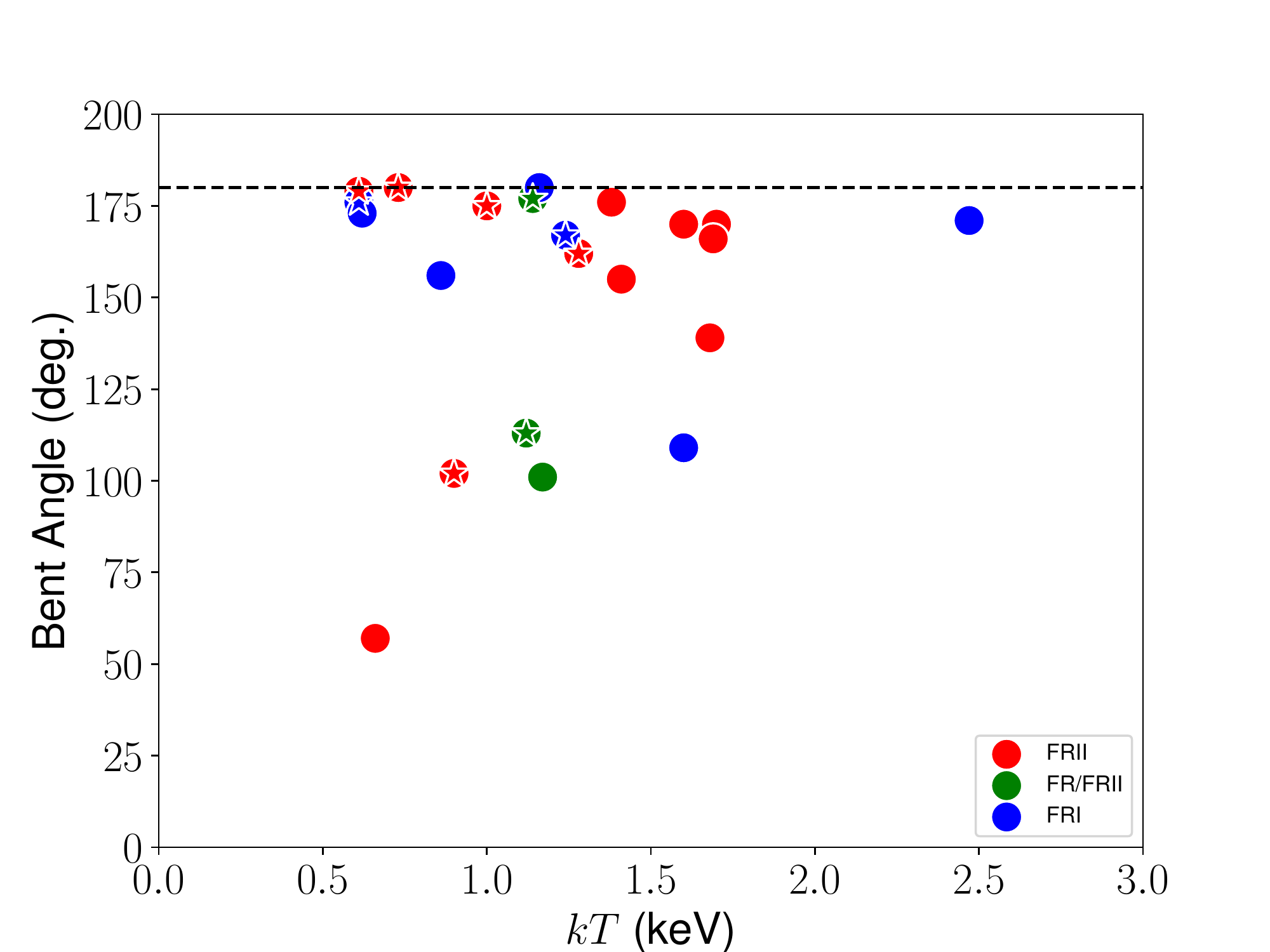}
            }
                 
       \caption{Bent angle in degrees for FRIIs (red), FRI/FRIIs (green), and FRIs (blue) with respect to the distance from the X-ray group centre normalised by the virial radius $r_{200}$ of the group (left), and with respect to the temperature $kT$ in keV of the X-ray galaxy group (right). The horizontal dashed line marks a straight radio source where the jets or lobes form a 180$^{\circ}$ angle.
   }
              \label{fig:bent_r200_kt}%
    \end{figure}

The analysis of BA and large-scale environments within COSMOS did not reveal any obvious correlation with BA. Thus we investigate possible correlations on smaller scales (kpc to 2 Mpc), within X-ray galaxy groups. We cross-correlate the FRs to the X-ray galaxy groups \citep{gozaliasl19} which cover the redshift range up to $z = 1.5$. In Fig.~\ref{fig:bent_r200_kt}--Left we plot the bent angle with respect to the distance from the X-ray galaxy group centre \citep{gozaliasl19}, normalised to the virial radius $r_{200}$ of each group. We mark the brightest galaxy in the group, BGG, with a star. We do not find a relation with the FR being associated with a BGGs and the location of the host with respect to the X-ray galaxy group centre. From the 13 FRs with straight radio structure (BA = 180 degrees), only 2 are members of X-ray galaxy groups.

For objects inside X-ray groups we have BA$_{\rm med}$ = 168.5 $\pm$ 35.9\,degrees, while outside the X-ray groups BA$_{\rm med}$ = 168.0 $\pm$ 17.8\,degrees. Beyond the redshift coverage of the X-ray galaxy groups, i.e. for $z >$ 1.5 we have BA$_{\rm med}$ = 167.0 $\pm$ 38.3 degrees. On average there is no difference in the BA of FRs inside X-ray galaxy groups and outside X-ray galaxy groups.

In Fig.~\ref{fig:bent_r200_kt}--Right we plot the bent angle with respect to the temperature $kT$ (0.5 $<$ keV $<$ 2) of the X-ray galaxy group \citep{gozaliasl19}. There is one FRI, source 773, with $kT \sim$ 2.5 keV. We do not find any strong correlation with bent angle and X-ray group temperature. This is probably related to the fact that X-ray galaxy groups in COSMOS are not yet dynamically relaxed, which was mentioned in the analysis of \citep{vardoulaki21} of FR radio source properties vs X-ray galaxy group properties.

\begin{figure}[!ht]
  \resizebox{\hsize}{!}
 {\includegraphics{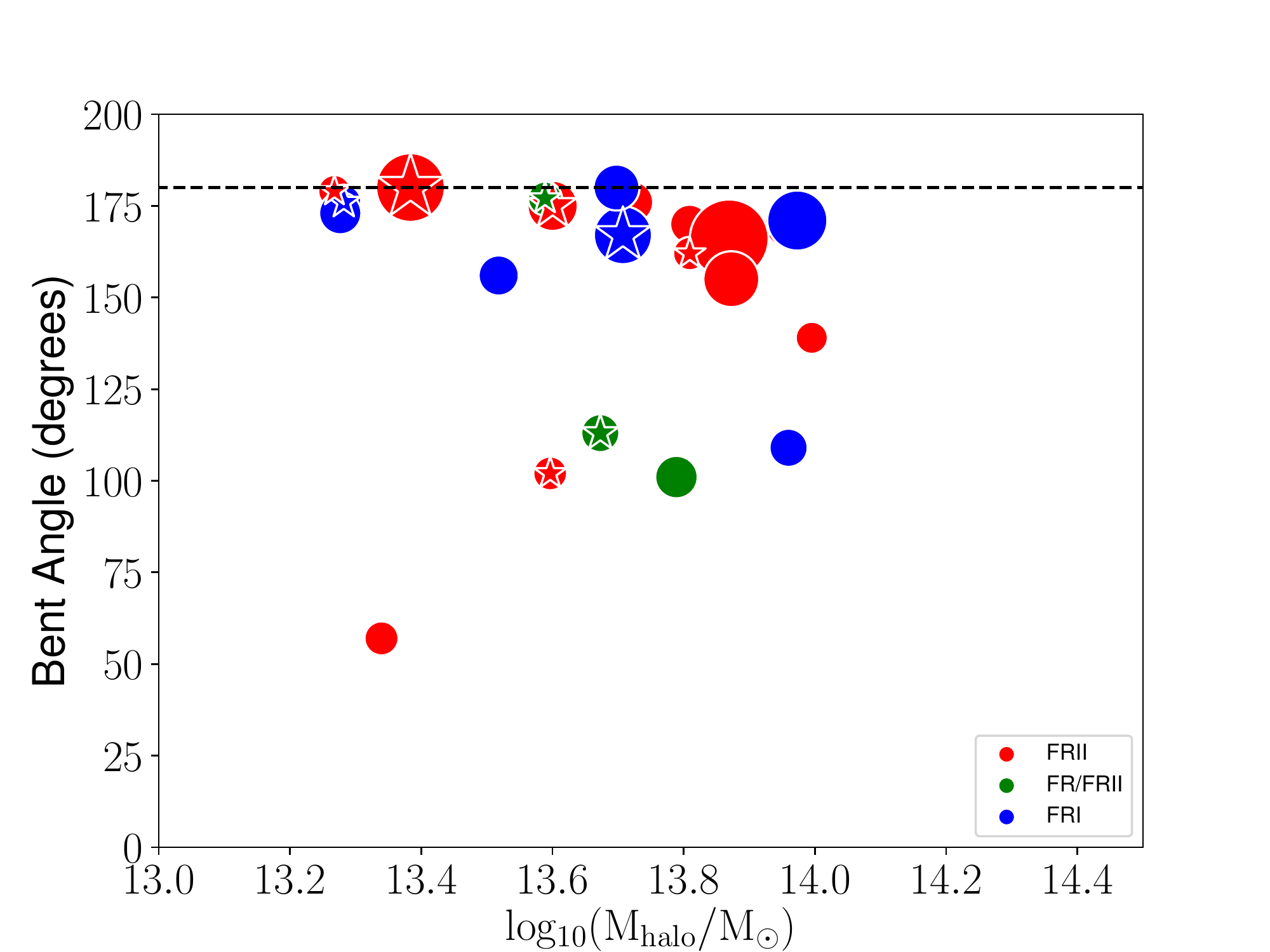}
 \includegraphics{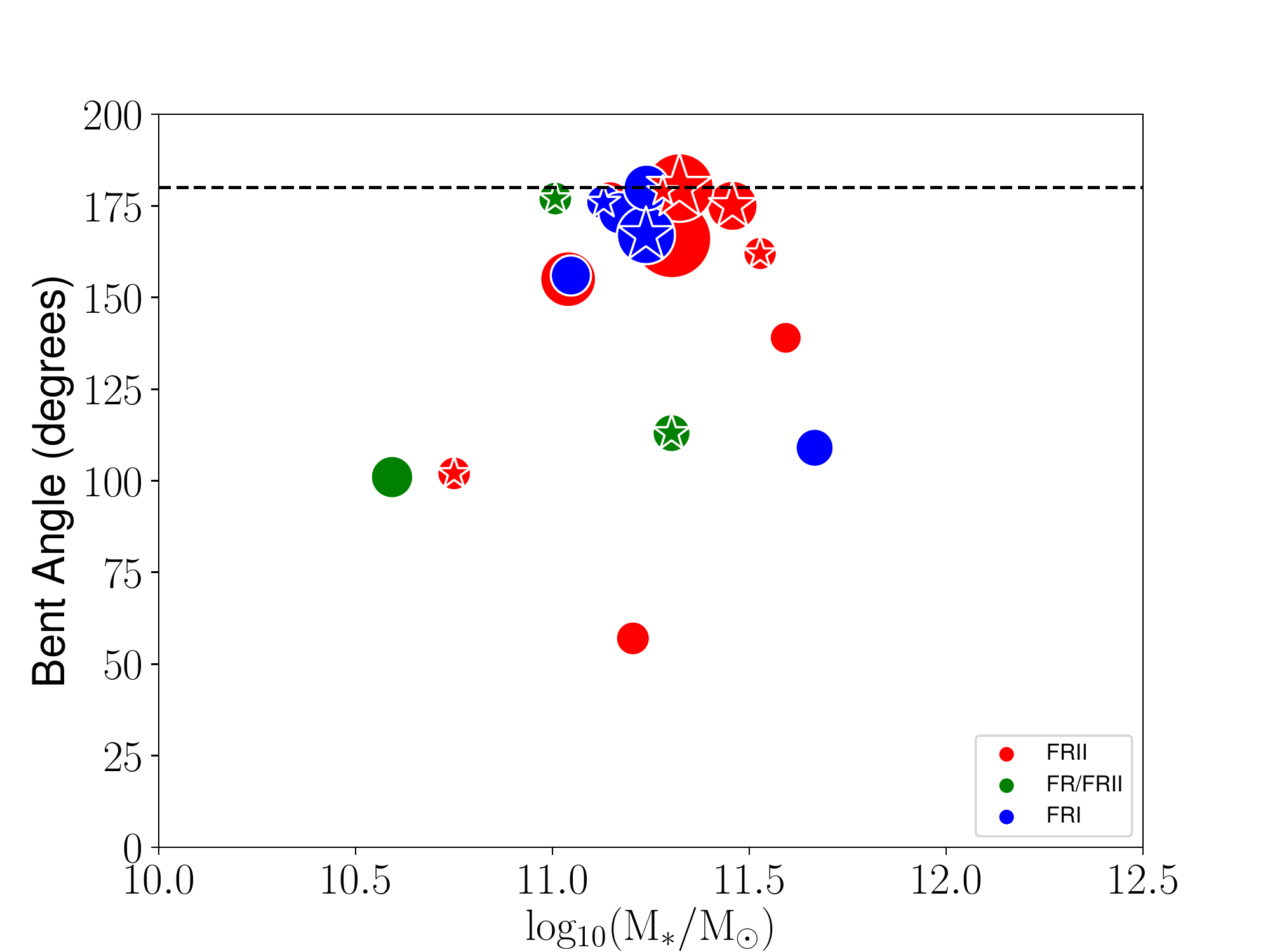}
            }
\caption{Bent angle in degrees for FRIIs (red), FRI/FRIIs (green), and FRIs (blue)  with respect to the mass $M_{200}$ of the X-ray galaxy group \citep{gozaliasl19, vardoulaki21b} at the left, and with respect to the stellar mass of the host $M_{*}$ at the right. The horizontal dashed line marks a straight radio source where the jets or lobes form 180$^{\circ}$. Symbol size is proportional to the $r/r_{200}$ ratio shown in Fig.~\ref{fig:bent_r200_kt}, where larger symbols show objects further away from the X-ray group centre. In all panels, stars denote the brightest group galaxy (BGG).
   }
              \label{fig:bent_mhalo}%
    \end{figure}

Additionally, in Fig.~\ref{fig:bent_mhalo} we compare the BA to the mass of the X-ray groups, $M_{200}$, and to the stellar mass of the host galaxy, $M_{*}$. There is no trend between X-ray group mass or host stellar mass with the BA. Furthermore, there is no significant difference in the BA of BGGs and non-BGGs regarding the mass of the group and the distance of the FR from the group centre. This can be seen in Fig.~\ref{fig:bent_mhalo} on the left panel, where the size symbol increases with distance from the X-ray group centre. Similarly with with the host stellar mass, on the right panel of Fig.~\ref{fig:bent_mhalo}.

Although at smaller scales, within galaxy groups, there are no obvious correlations between BA and group properties, we find that the BA is always small (close to 180 degrees) for sources with large distance to the X-ray group centre and high temperature of the X-ray group. The BA of radio galaxies closer to the X-ray group centre and with low temperature shows a large scatter.

\subsection{Comparison to simulations}

Vazza et al. \citep{vazza21} simulated the evolution of relativistic electrons injected into the ICM by radio galaxies, with a simplistic model in which each radio galaxy only releases radio lobes once during its evolution. They perform MHD adaptive-mesh simulations with ENZO (enzo-project.org) using passive tracer particles to track the propagation of cosmic ray electrons. By incorporating energy losses and re-acceleration mechanisms they calculate the energy evolution of the spectrum of radio emitting electrons as they propagate and get dispersed into the ICM. 

\begin{figure}[!ht]
  \resizebox{\hsize}{!}
 {\includegraphics{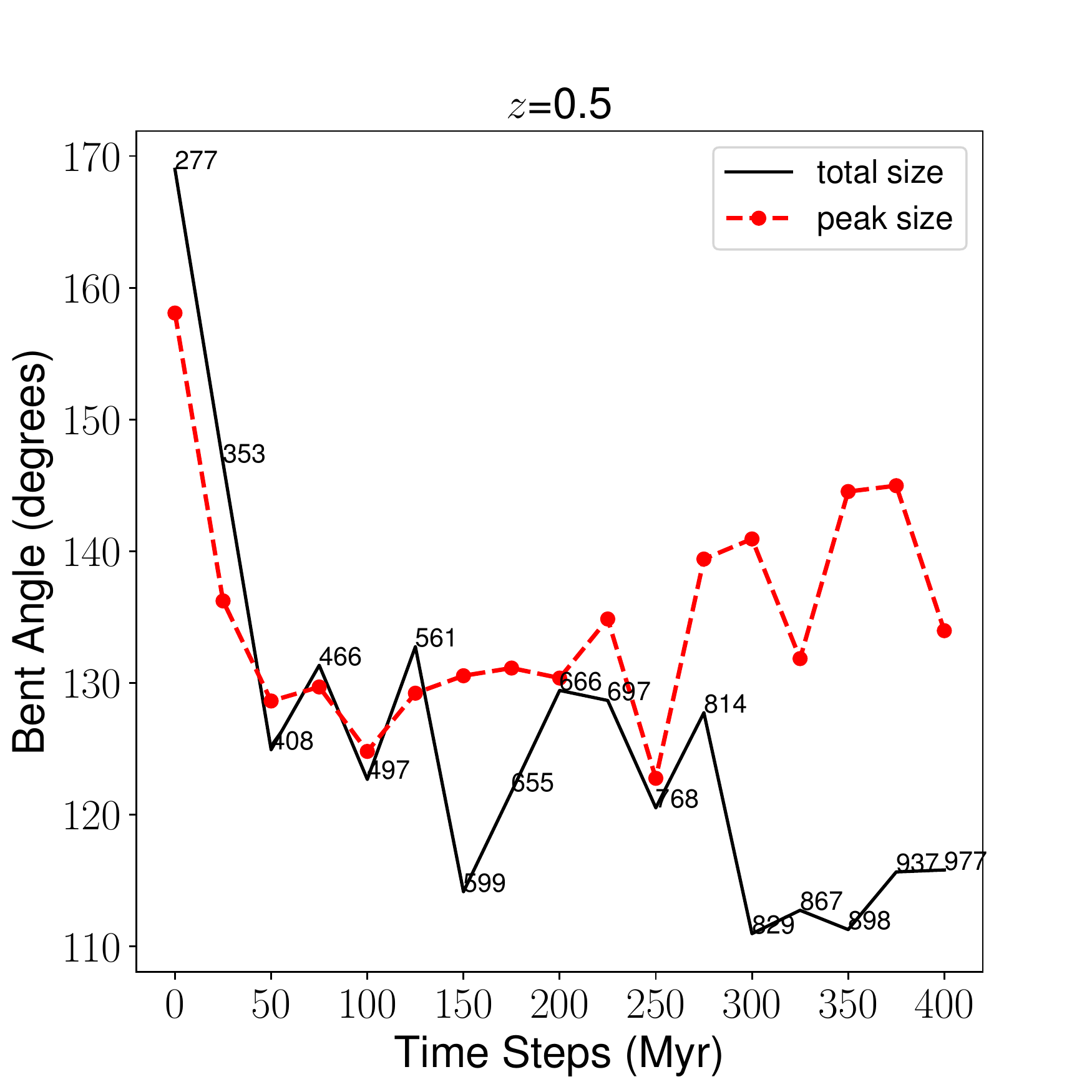}
 \includegraphics{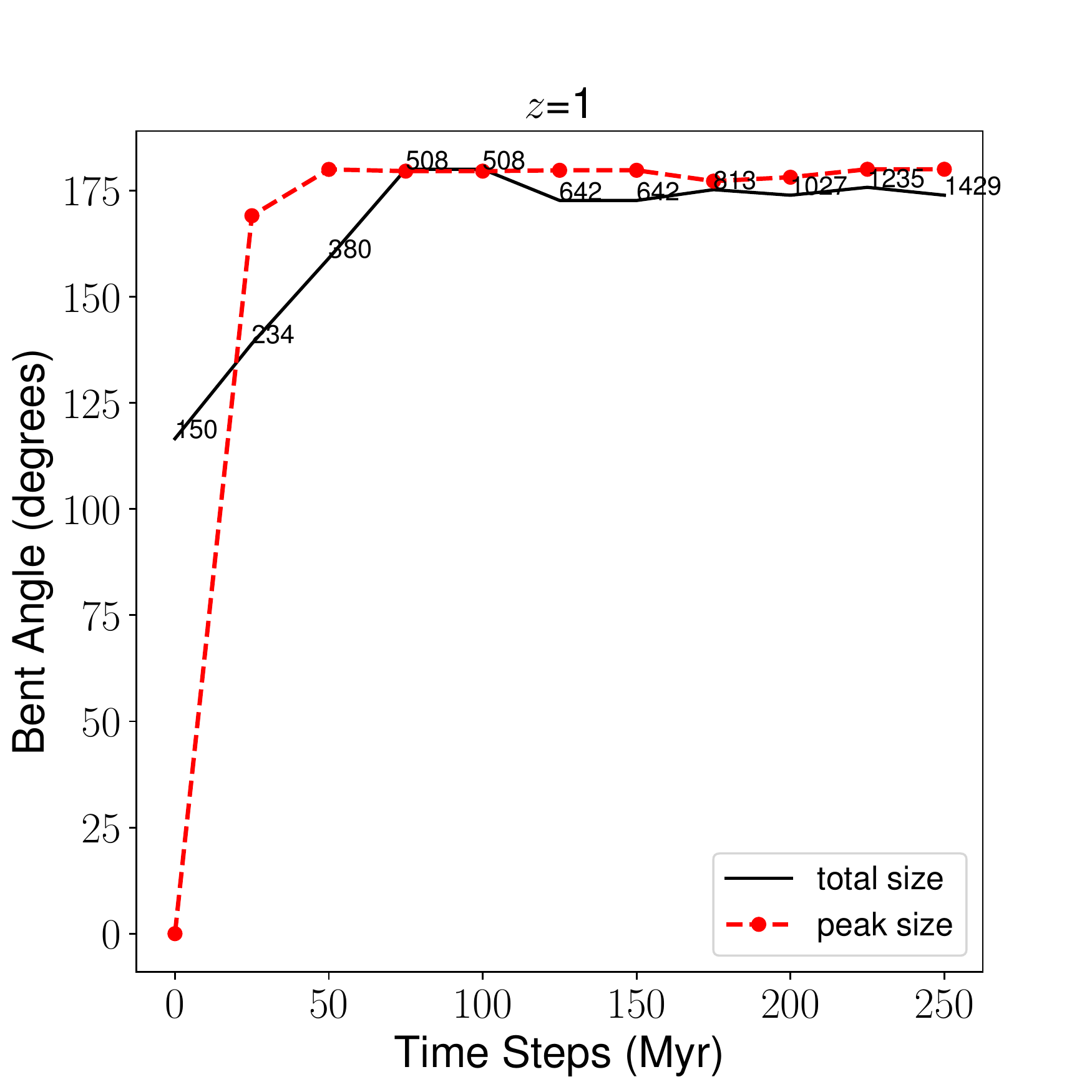}}

       \caption{Simulated radio AGN at 1.4 GHz from \cite{vazza21} at $z$ = 0.5 (left) and $z$ = 1 (right). We traced the evolution of the bent angle formed between source edges (black solid line) and source peak flux (red dotted line) at both ends every 25 Myrs. The numbers show the linear projected size of the source in kpc.  
   }
              \label{fig:bent_sim_vaz}%
    \end{figure}

The two simulated radio galaxies are at redshifts 0.5 and 1. Here we analysed the evolution of each simulated source, how it grows in size every 25 Myrs, in relation to the angle formed between the lobes/jets. We traced the position of the highest flux peak in the lobe compared to the core position. This is shown as red dotted lines in Fig.~\ref{fig:bent_sim_vaz}. We also trace the BA formed between the two edges of the sources, shown by the black solid line. For each 25 Myr time step we mark how the linear projected size of the source increases given time for reference  \footnote{Movies of the two simulations can be found at \hreflink{https://vimeo.com/490397871}{https://vimeo.com/490397871} and \hreflink{https://vimeo.com/490399056}{https://vimeo.com/490399056}.}.
  
Comparison of our COSMOS data from Fig.~\ref{fig:bent_sim_vaz} to the ENZO-MHD simulated sources at $z$ = 0.5 and at $z$ = 1 suggests a good agreement within the scatter (see Table~\ref{tab:ba_simul}). The median BA of our COSMOS objects at a median redshift of around one is 167.5 degrees. The most bent objects in our sample are at lowest redshifts. To compare our observational results to the simulations we calculate the median BA at two redshift bins, $0.4 < z < 0.6$ and $0.9 < z < 1.1$, shown in Table~\ref{tab:ba_simul}. For the $z \sim$ 1 redshift bin we find the observations (BA$_{\rm med}$ =  174.0 $\pm$ 44.9) and simulations (BA$_{\rm med}$ =  173.8 $\pm$ 52.4) agree, within the scatter in the values. For the $z \sim$ 0.5 redshift bin, the observations (BA$_{\rm med}$ =  160.0 $\pm$ 53.3) give higher on average bent angle than the simulated source (BA$_{\rm med}$ = 122.6 $\pm$ 29.7), but the median BA for the simulated source is within the scatter of the observations. This discrepancy suggests some sources of uncertainty in the observations. These could be projection effects more pronounced at low redshifts, classification biases, or different different density environments. Based on our analysis of BA vs density fields, FRIIs and FRI/FRIIs lie, on average, in denser environments at $z \sim$ 0.5 than at $z \sim$ 1 (by a factor of 2). The FRIs are found, on average, in less dense environments throughout. The difference between the environments of FRI and FRII, FRI/FRII could justify the large scatter in BA.

For objects in COSMOS below redshift of 0.5, we have a median BA = 167.5 and error of the median $\pm$ 17.5 degrees, which also falls in the range of values of the simulate source (Table~\ref{tab:ba_simul}). For objects between $0.5 < z < 1.5$ we get a median BA of 167.0 $\pm$ 22.2 degrees. Finally, objects above redshift of 1.5, beyond what our large-scale environments in COSMOS cover (apart from the density fields) and beyond the Vazza et al. simulations, have median BA = 168.0 $\pm$ 37.5 degrees. The difference in bent angle between the low- and high-$z$ simulated sources could be attributed to differences in the IGM density and the structure of the IGM itself. The median density value, as extracted from the simulation is $1.19 \pm 0.28 10^{-27} g/cm^{3}$ at $z$ = 0.5 and $1.73 \pm 0.52 10^{-28} g/cm^{3}$ at $z$ = 1, an order of magnitude less. Additionally, the IGM at lower $z$ has a longer depth, meaning it is more extended. In other words, the lower the redshift the wider would be the IGM distribution, which can promote more interaction and bending of the lobes.

\begin{specialtable}[H] 
\caption{Median bent angle BA in different redshift bins within COSMOS and comparison to the ENZO-MHD simulated sources \cite{vazza21}. The BA of the simulated sources is traced over time.}
\label{tab:ba_simul}
%%% \tablesize{} %% You can specify the fontsize here, e.g., \tablesize{\footnotesize}. If commented out \small will be used.
\begin{tabular}{l l l | l l l}
\toprule
 \multicolumn{3}{c}{\textbf{Observations}} & \multicolumn{3}{c}{\textbf{Simulation}} \\
\multicolumn{1}{c}{\textbf{Redshift ($z$)}} & 
\multicolumn{1}{c}{\textbf{$N$}} &
\multicolumn{1}{c}{\textbf{Bent angle$^{84\%}_{16\%}$ (deg.)}} &
\multicolumn{1}{c}{\textbf{$N$}} &
\multicolumn{1}{c}{\textbf{$z_{\rm sim}$}} & \multicolumn{1}{c}{\textbf{Bent angle$^{84\%}_{16\%}$ (deg.)}}  \\     
\midrule
$0.4 < z < 0.6$ & 9 & 160.0$^{178.6}_{111.5}$  & 0.5 & 1 & 122.6$^{131.9}_{113.5}$\\
& & &\\
$0.9 < z < 1.1$ & 15 & 174.0$^{178.2}_{163.9}$  & 1 & 1 & 173.8$^{177.4}_{151.0}$ \\
& & &\\
\hline
& & &\\
$z_{\rm med}~ \sim$  0.9 & 108 &167.5$^{179.0}_{130.0}$ \\
& & &\\ 
$z ~<$ 0.5 & 88 &  167.5$^{179.0}_{112.8}$ & \\
& & &\\
$0.5 ~< z ~< 1.5 $ & 58 &169.5$^{179.0}_{141.8}$ &\\
& & &\\
$ 1.5 ~< z ~< 3$ & 20 &168.0$^{175.0}_{139.5}$ \\ 
& & &\\
\bottomrule
\end{tabular}
\end{specialtable}

%%%%%%%%%%%%%%%%%%%%%%%%%%%%%%%%%%%%%%%%%%
\section{Discussion}

Our analysis of the projected bent angle of 3 GHz COSMOS FR-type radio sources gives a wide range of values, from very bent sources with BA $\sim$ 40 degrees to almost straight double-sided FR radio structures. In the context of galaxy evolution and AGN nature, we find it interesting that there are no obvious correlations with the large-scale environment and bent angle of extended radio AGN. 

Within X-ray galaxy group environments (Fig.~\ref{fig:bent_r200_kt}) we find only 24 out of the 75 FRs, at the same redshift range and sky area coverage. This finding is related to the current limitations of X-ray surveys to detect galaxy groups with masses below $1.5 (1 + z) \times 10^{13}M_{\odot}$ \cite{vardoulaki19}, and bent FRs outside galaxy groups can be used to identify these. We only find 2 out of the 13 (15\%) straight FRs (BA = 180 degrees) inside the X-ray groups, which could be either related to the limitations of the current X-ray observations in COSMOS, or straight FRs do not prefer X-ray group environments. It could also mean the radio sources which are straight did not have time to interact with the medium and get their jets bent. Improvement over the COSMOS X-ray observations can shed light to this. 

Regarding the density environments, most of the bent FRs in our sample are found at densities of 1/Mpc$^{2}$ (Fig.~\ref{fig:bent_dens}), with a handful above densities of 10/Mpc$^{2}$. The mild relation we find between FR type and BA is interesting and puzzling at the same time, given the FRIs, which are on average slightly more bent than FRIIs, are found in less dense environments than the other FR-types at 3 GHz VLA-COSMOS. The latter result is also reported in \cite{vardoulaki21}, showing that FRIs reside in regions where environmental density, the number of galaxies per Mpc$^{2}$, is lower than the other FR types. The smaller bent angles of FRIs could be caused by either movement of the jets through a denser ambient medium, or movement of the objects through the ICM, in which case ram pressure bents the jets backwards  \citep[e.g.][]{smolcic07}. Given the large scatter of BA values in the FRI sample, and given the average environmental density of FRIs is lower than FRIIs with an overlap in the distributions, we believe this is an object-to-object investigation. The bent angle is likely affected by more than one parameter.

The statistics given the cosmic web probes (Fig.~\ref{fig:bent_darv}) are small and the uncertainty too large for a robust result. There is a mild trend for more bent FRs to lie in filaments. FRIs associated with satellite hosts are more bent than FRI/FRIIs and FRIIs. The same is seen if the host is a central galaxy, but the number of objects is small and the scatter large. Bent radio sources located several Mpc from cluster centres, in filaments, have been reported before \citep{edwards10} and can be used to probe inter-filament density ($\sim 10^{-4}-10^{-6}$ cm$^{-3}$; see also \cite{plank13}), which is less dense than what is expected for the gas in clusters. We understand that the interplay between hosts associated with FRIs plays a role in shaping their radio structures. If these hosts are satellite galaxies, which are less dynamically stable, it would explain the reason for distortions in the radio structure.

To investigate how the BA values of the FRs in our sample compare to other well-studied samples, we take the radio galaxy zoo project \citep[RGZ;][]{garon19}, which studied 4304 bent radio galaxies inside clusters. They found that within the $\sim$ 1 Mpc of the cluster, galaxies not associated with the brightest cluster galaxy, non-BCGs,  are statistically more bent in high-mass clusters than in low-mass clusters. Contrary to that, we do not see a difference in the bent angle of non-BGG in relation to the group mass. Additionally, we do not find a difference between non-BGGs and BGGs in the inner 1 Mpc (Fig.~\ref{fig:bent_mhalo}). Furthermore, we do not find a difference in BA between non-BGGs and BGGs as we move closer to the group centre (see Fig.~\ref{fig:bent_r200_kt}). This can also be seen in the left panel of Fig.~\ref{fig:bent_mhalo} where the size of the symbol increases with the distance from the group centre.

The origin of such discrepancies between our study and the RGZ study might be a result of different selection criteria. Firstly, the total number of sources in our sample is $\sim$ 2\% of that of the RGZ sample. Secondly, while our sample includes galaxy group masses in the range $M_{\rm 200c} = 8 \times 10^{12} - 3 \times 10^{14} M_{\odot}$, the RGZ sample includes cluster masses $M_{\rm 500c} = 5 - 30 \times 10^{14} M_{\odot}$. To compare the COSMOS galaxy groups halo masses to the RGZ cluster halo masses we calculated the concentration and converted $M_{\rm 200c}$ to $M_{\rm 500c}$ using the code Colossus \citep{diemer18}. The parameters we used in their code are: Planck15 cosmology and median redshift of COSMOS of 0.9. We obtained $M_{\rm 500c} = 5 \times 10^{12} - 2 \times 10^{14} M_{\odot}$. COSMOS is probing galaxy groups with smaller halo masses than probed by RGZ, and these groups are not dynamically relaxed as yet \cite{vardoulaki21}. This could explain the difference in the results between the two studies regarding the BA of BGGs and non-BGGs.

Our observational result that lower redshift FRs are, on average, more bent than higher redshift FRs, suggests their energy deposit on the ambient medium is not enough to sustain the straight radio structure. Although our study is affected in places by small number statistics, and possible projection effects, our results are in line with the ENZO-MHD simulations \citep{vazza21} presented here. Thus, there seems to be another important parameter which defines the median BA of radio sources, revealed by the comparison to the simulations: the redshift evolution of the IGM, which is well reproduced by the observational data. The IGM density at lower redshift is higher and has longer depth than at the higher redshift, which leaves more room for jet interactions. Thus the radio jets get more perturbed and bent as they move across the IGM.

From the comparison to the simulations \citep{vazza21} we infer that the setup of the ENZO-MHD simulations can mimic real sources. This is more obvious at the higher redshift bin than the lower, when compared to our sample of COSMOS FRs. What we see in the FRs in COSMOS is a snapshot of their evolution. The large scatter in our data compared to the simulated values represents the complexity of studying these objects in relation to their environment, as well as how difficult it is to understand the reason for the bent radio structure. We also see them at different stages of their evolution. We believe this is a case to case problem and needs to be approached in a multi-parameter manner. This will be the topic of a future study.

%%%%%%%%%%%%%%%%%%%%%%%%%%%%%%%%%%%%%%%%%%
\section{Conclusions}

We presented the analysis of the bent radio structure of FR-type radio AGN from the COSMOS field using the 3 GHz VLA-COSMOS data \citep{smolcic17a,vardoulaki21}. To parametrise the complexity we use measurements of the bent angle \citep{vardoulaki21}, the angle the highest flux density peaks in the lobes/jets form to each other. Our sample of 108 FRs has a median BA of 167.5$^{+11.5}_{-37.5}$ degrees at a median redshift of 0.9. We investigate the relation of the BA with several large-scale environmental probes within COSMOS (X-ray galaxy groups, density fields, and cosmic web). This study covers, for the first time, galaxy group halo masses $M_{\rm 200c} = 8 \times 10^{12} - 3 \times 10^{14} M_{\odot}$, and the redshift range 0.08 $< z < 3$ and scales from a few kpc up to 500 Mpc. Finally, we compare our observational BA values to the ones derived from the ENZO-MHD simulations of two radio sources at $z = 0.5$ and at $z = 1$ \cite{vazza21}.

 Our results are summarised as follows:
 \begin{enumerate}
     \item We find a mild correlation between BA and FR type.
     \item We do not find a correlation between BA and large-scale environments or galaxy type.
     \item We do not find a significant correlation with BA and X-ray galaxy group properties. More straight sources with higher group temperatures lie further away from the group centre, while the BA of radio galaxies closer to the X-ray group centre and with low temperature shows a large scatter.
     \item The difference with large studies such as RGZ can be attributed to the fact that COSMOS is probing lower mass halos and galaxy groups which are not dynamically relaxed as yet.
     \item FRI are, on average, slightly more bent than FRII, and lie in less dense environments.
     \item The redshift evolution of BA in the simulations from $z = 1$ to $z = 0.5$ is reproduced by the observational data, within the scatter.
 \end{enumerate}

 Investigating relations to the large-scale environment is interesting in the COSMOS field and not intuitive given the fact we observe the faint radio universe at high sensitivity (2.3 $\mu$Jy/beam) and high resolution (0".75). We do not find significant trends with large-scale environments within COSMOS. Comparison to simulations suggests a link between BA and redshift evolution. More bent sources at lower redshift can be explained with higher density and larger depth of the ambient medium, giving space to FRs to interact more. The larger scatter in BA at lower redshifts in our observational data could be explained by the difference between FRI and FRII, FRI/FRII environments, as probed by the density fields in COSMOS. 
 
 We conclude that the dominant mechanism affecting the radio structures of FRs could be the IGM evolution with redshift, with higher density and longer depth at lower redshifts, allowing for more jet interactions, and not the large-scale environment. Our study, although limited in places due to small number statistics,  demonstrates the benefit of multi-wavelength data at a very well studied field like the COSMOS field. Future radio surveys in the SKA era, with increased sensitivity and resolution beyond current capabilities, can benefit from studies like this one and perform better statistics on extended radio AGN in order to identify the reasons behind distorted radio structures. Nevertheless, we believe that the parameter that is responsible for bent radio AGN is not only one but can be a combination of parameters, from smaller scales within the host galaxy to larger scales outside the host, which we will investigate with future studies. 

%\begin{enumerate}
%      \item 
%   \end{enumerate}

%%%%%%%%%%%%%%%%%%%%%%%%%%%%%%%%%%%%%%%%%%
\vspace{6pt}

%%%%%%%%%%%%%%%%%%%%%%%%%%%%%%%%%%%%%%%%%%

\acknowledgments{This study was presented by EV at the "A new window on the radio emission from galaxies, clusters and cosmic web" virtual conference on 10 March 2021, during the Covid-19 pandemic. EV acknowledges support by the Carl Zeiss Stiftung with the project code KODAR. The authors gratefully acknowledge the Gauss Centre for Supercomputing e.V. (www.gauss-centre.eu) for supporting this project by providing computing time through the John von Neumann Institute for Computing (NIC) on the GCS Supercomputer JUWELS at J\"ulich Supercomputing Centre (JSC), under project  "radgalicm". F.V. acknowledges financial support from the European Union's Horizon 2020 program under the ERC Starting Grant "MAGCOW", no. 714196. D.W. is funded by the Deutsche Forschungsgemeinschaft (DFG, German Research Foundation) - 441694982.}

%%%%%%%%%%%%%%%%%%%%%%%%%%%%%%%%%%%%%%%%%%

\reftitle{References}


%\externalbibliography{yes}
%\bibliography{your_external_BibTeX_file}

%=====================================
% References, variant B: internal bibliography
%=====================================

% If authors have biography, please use the format below
\section*{Short Biography of Authors}
\bio
{\raisebox{-0.35cm}{\includegraphics[width=3.5cm,height=5.3cm,clip,keepaspectratio]{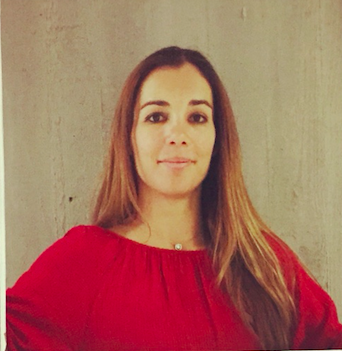}}}
{\textbf{Eleni Vardoulaki} is coordinator for data-intensive radio astronomy and a postdoctoral researcher at Th\"{u}ringer Landessternwarte Tautenburg (TLS), and visitor researcher at Max-Planck-Institut f\"{u}r Radioastronomie (MPIfR) Bonn. DPhil from the University of Oxford (2010). Postdoc researcher at CAUP (Porto, Portugal), University of Crete (Greece), Argelander-Institut f\"{u}r Astronomie Bonn and MPIfR. Multi-wavelength studies of radio sources, AGN/SFG disentangling, high and low redshift universe, machine learning. Member of the American, European, and Hellenic Astronomical Societies. Member of the COSMOS, EMU, MeerKAT-MIGHTEE collaborations. TEDx speaker and science communicator (Rogue Astrophysics; Astronomy on Tap; elenivardoulaki.com)}
%\bio
%{\raisebox{-0.35cm}{\includegraphics[width=3.5cm,height=5.3cm,clip,keepaspectratio]{Definitions/author2.jpg}}}
%{\textbf{Firstname Lastname} Biography of second author}

% The following MDPI journals use author-date citation: Arts, Econometrics, Economies, Genealogy, Humanities, IJFS, JRFM, Laws, Religions, Risks, Social Sciences. For those journals, please follow the formatting guidelines on http://www.mdpi.com/authors/references
% To cite two works by the same author: \citeauthor{ref-journal-1a} (\citeyear{ref-journal-1a}, \citeyear{ref-journal-1b}). This produces: Whittaker (1967, 1975)
% To cite two works by the same author with specific pages: \citeauthor{ref-journal-3a} (\citeyear{ref-journal-3a}, p. 328; \citeyear{ref-journal-3b}, p.475). This produces: Wong (1999, p. 328; 2000, p. 475)

%%%%%%%%%%%%%%%%%%%%%%%%%%%%%%%%%%%%%%%%%%
%% for journal Sci
%\reviewreports{\\
%Reviewer 1 comments and authors’ response\\
%Reviewer 2 comments and authors’ response\\
%Reviewer 3 comments and authors’ response
%}
%%%%%%%%%%%%%%%%%%%%%%%%%%%%%%%%%%%%%%%%%%
\end{paracol}
\end{document}